\documentstyle [12pt] {article}

\newcommand{\gtwid}{\mathrel{\raise.3ex\hbox{$>$\kern-.75em\lower1ex
\hbox{$\sim$}}}}
\newcommand{\ltwid}{\mathrel{\raise.3ex\hbox{$<$\kern-.75em\lower1ex
\hbox{$\sim$}}}}
\newcommand{\beq}{\begin{equation}}
\newcommand{\eeq}{\end{equation}}
\newcommand{\beqs}{\begin{eqnarray}}
\newcommand{\eeqs}{\end{eqnarray}}

\catcode`@=11
\@addtoreset{equation}{section}
\@addtoreset{equation}{subsection}
\def\theequation{\ifnum\value{section}=0 \arabic{equation}\ignorespaces
\else \ifnum\value{section}=-1 A.\arabic{equation}\ignorespaces
\else \ifnum\value{subsection}=0 \thesection.\arabic{equation}\ignorespaces
\else \thesection.\arabic{subsection}.\arabic{equation}\ignorespaces
                           \fi
                      \fi
                 \fi}
\catcode`@=12

\begin{document}

\def\thefootnote{\fnsymbol{footnote}}
\baselineskip 6.0mm

\begin{flushright}
\begin{tabular}{l}
ITP-SB-95-9    \\
April, 1995
\end{tabular}
\end{flushright}

\vspace{8mm}
\begin{center}

{\Large \bf Zeros of the Partition Function for }

\vspace{2mm}

{\Large \bf Higher--Spin 2D Ising Models}

\vspace{16mm}

\setcounter{footnote}{0}
Victor Matveev\footnote{email: vmatveev@max.physics.sunysb.edu}
\setcounter{footnote}{6}
and Robert Shrock\footnote{email: shrock@max.physics.sunysb.edu}

\vspace{6mm}
Institute for Theoretical Physics  \\
State University of New York       \\
Stony Brook, N. Y. 11794-3840  \\

\vspace{16mm}

{\bf Abstract}
\end{center}
    We present calculations of the complex-temperature zeros of the partition
functions for 2D Ising models on the square lattice with spin $s=1$,
3/2, and 2.  These give insight into complex-temperature phase diagrams of
these models in the thermodynamic limit.  Support is adduced for a conjecture
that all divergences of the magnetisation occur at endpoints of arcs of zeros
protruding into the FM phase.  We conjecture that there are $4[s^2]-2$ such
arcs for $s \ge 1$, where $[x]$ denotes the integral part of $x$.
\vspace{16mm}

\pagestyle{empty}
\newpage

\pagestyle{plain}
\pagenumbering{arabic}
\renewcommand{\thefootnote}{\arabic{footnote}}
\setcounter{footnote}{0}

  The Ising model has long served as a simple prototype of a
statistical mechanical system which (for spatial dimensionality $d > 1$)
undergoes a phase transition with associated spontaneous symmetry breaking
(SSB)
and long range order.  The zero-field $d=2$ spin 1/2 Ising model is exactly
solvable; the free energy $f$ and spontaneous magnetisation $M$ were first
derived by Onsager and Yang, respectively \cite{ons,y} (both for the square
lattice).  However, no exact closed--form expressions have ever been found
for $f$ or any other thermodynamic quantity for any higher--spin Ising model
in $d=2$ (or higher $d$).  It is therefore of considerable value to establish
further properties of higher--spin Ising models, since these help one to
make progress toward an exact solution.  In this letter we present
calculations of the complex-temperature (CT) zeros of the partition functions
of the 2D Ising model on the square lattice for the higher spin values
$s=1$, 3/2, and 2.  These results enable one to infer information about the CT
phase diagrams in the thermodynamic limit for these models.
Combining our results with analyses of low-temperature series by Guttmann et
al. \cite{egj,jge}, we adduce support for our conjecture on CT divergences in
$M$.  We also infer a new conjecture on arcs of zeros.

   There are several reasons for interest in CT properties of spin models:
(i) one can understand more deeply the behaviour of various thermodynamic
quantities by seeing how they behave as analytic functions of CT (e.g., CT
singularities can affect behaviour for physical temperature); (ii) one can
see how the physical phases of a given model generalise to regions for CT;
(iii) as noted, a knowledge of the CT properties helps in the search for exact
solutions. The earliest papers on CT properties in the Ising model include
Refs. \cite{fisher}-\cite{g75} for $s=1/2$; CT singularities for higher--spin
(2D and 3D) Ising models were first studied in Ref. \cite {fg} (see also
Refs. \cite{fg1,sg}). Later papers for $s=1/2$ include Refs.
\cite{ipz}-\cite{cmo}.

    The spin $s$ (nearest-neighbour) Ising model on the square lattice is
defined, in standard notation, by the partition function
$Z = \sum_{\{S_n\}} e^{-\beta {\cal H}}$, with
\beq
{\cal H} = -(J/s^2) \sum_{<nn'>} S_n S_{n'} - (H/s) \sum_n S_n
\label{ham}
\eeq
where $S_n \in \{-s, -s+1,...,s-1,s\}$ and $\beta = (k_BT)^{-1}$. $H=0$
unless otherwise indicated. We define $K = \beta J$,
$v = \tanh K$, and $u_s = e^{-K/s^2}$.  $Z$ is then a generalised (i.e. with
negative as well as positive powers) polynomial in $u_s$.  We denote
the physical critical point as $K_c$ for a given $s$, and
$(u_s)_c = \exp(-K_c/s^2)$.  The (reduced) free energy is
$f = -\beta F = \lim_{N_s \to \infty} N_s^{-1} \ln Z$ in the thermodynamic
limit.

   Early series studies (mainly for 3D) gave evidence that
the critical exponents of the phase transition between the $Z_2$--symmetric,
paramagnetic (PM) phase and the phase with SSB and long range ferromagnetic
(FM) order are independent of $s$ \cite{fg1,crit}.
This is in agreement with expectations from renormalisation group arguments,
since changing $s$ does not change the $Z_2$ symmetry group of ${\cal H}$.
However, for CT, the value of $s$ does have interesting effects.

    Since there is no closed-form expression for $f$ for $s \ge 1$, one
cannot directly determine the locus of CT points where $f$ is non-analytic
(aside from the points $K=\pm \infty$ where it is trivially non-analytic).
However, from calculations of zeros for the 2D spin 1/2 Ising model and
comparison with exact results, one knows that as the lattice size increases,
the zeros of $Z$ occur on, or progressively closer to, the above locus of
points where, in the thermodynamic limit, $f$ is non-analytic.
By studying the zeros
of $Z$ on finite lattices with varying boundary conditions (BC), one can thus
make reasonable inferences about this locus of points and the corresponding CT
phase diagram.

   For the $d=2$, $s=1/2$ Ising model with isotropic couplings $J$
on the square lattice (as well as on the triangular
and honeycomb lattices, and also certain heteropolygonal lattices \cite{cmo}),
as $N_s \to \infty$, these zeros of the partition function in
$z$ merge to form continuous curves, including possible line
segments.  The CT phase diagram is comprised of phases bounded by
such curves where leading and next-to-leading eigenvalues of the transfer
matrix ${\cal T}$ (which are, in general, complex here) become degenerate in
magnitude and hence there is a non-analytic change in $f$. We show now that
(for isotropic $J$) the zeros of $Z$ again merge to form a one-dimensional
variety (i.e., curves including possible line segments, instead of
areas) in the complex $u_s$ plane for arbitrary $s$.  Using the the discrete
translation invariance of the square lattice, one can, by Fourier
transformation methods, write
\beq
f= \ln (2s+1) + \int_{-\pi}^{\pi}\int_{-\pi}^{\pi} \frac{d\theta_1
d\theta_2}{(2\pi)^2} g \Bigl ( A + B(\cos \theta_1 + \cos \theta_2) \Bigr )
\label{f}
\eeq
where $A$, $B$ are functions of $u_s$. The locus of points where $g$ and hence
$f$ are non-analytic is given by an equation involving only the argument of
$g$, namely, $A + B(\cos \theta_1 + \cos \theta_2)$.  Now since the two
independent (periodic) variables $\theta_1$ and $\theta_2$ only enter in the
combination $\cos \theta_1 + \cos \theta_2 \equiv x$, the argument of $g$ can
be written as
$A(u_s) + B(u_s)x$, so that the equation involves a single independent real
variable ($-2 \le x \le 2$).  Hence, the continuous locus of points where $f$
is non-analytic (i.e., excluding the trivial isolated isolated points
$K = \pm \infty$) is a one-dimensional variety in the $u_s$ plane.
The zeros of $Z$ and corresponding curves
where $f$ is non-analytic are invariant under $u_s \to u_s^*$ and also, for a
bipartite lattice, under the mapping $u_s \to 1/u_s$.

   In order to compute the zeros, we calculate $Z$
for finite lattices with specified BC's.  We have done this by means of a
transfer matrix method \cite{tm}, and have used both
periodic and helical BC's (resp. PBC, HBC). We have also used a
type of helical BC introduced by Creutz (CHBC) \cite{mike} which reduces
finite-size effects.  We follow the notation of Ref. \cite{mike} to label
the lattice in this case: $N_1 \times (N_s/N_1)$, where $N_s$ denotes the
total number of sites and $N_1$ denotes the number of
sites in all the rows except the last.

    In Fig. 1 we show the zeros of $Z$
for the Ising model on a square lattice with (i) $s=1$, lattice size
$8 \times 10$, PBC; (ii) $s=3/2$, lattice size $6 \times 6$, PBC; (iii)
$s=2$, lattice $5 \times (38/5)$, CHBC.  The horizontal and vertical axes are
$Re(u_s)$ and $Im(u_s)$ for each $s$. We have calculated
zeros for a number of different lattices.  Although we only show one plot for
each $s$ here, the features which we point out were observed on all of the
lattices used, with both PBC's and (C)HBC's.

   First, up to some slight scatter, the zeros can be connected by curves, in
agreement with our general argument above. Second, for each $s$, the model
has physical PM, FM, and AFM phases, and these have analytic continuations to
form CT phases in the $u_s$ plane;  we denote these as (CT) PM, FM, and AFM
phases.  The symmetry of the phases about the unit circle $|u_s|=1$ follows
from the $u_s \to 1/u_s$ symmetry of the model.

   Third, the CT phase diagram also includes phases (labelled ``O'') which
have no overlap with any physical phase.  For $s=1,2$ there is an O$_1$ phase
to the left of the FM phase separating it from the AFM phase.  We conjecture
that this is true for abitrary integral $s$. We restrict this conjecture to
integral $s$ since we already know that it does not hold in the exactly solved
$s=1/2$ model; in that case, there is no O phase (in the $u_{1/2}
\equiv e^{-4K}$ variable), and the FM and AFM (and PM) phases are directly
contiguous at $u_{1/2}=-1$ (see Fig. 1(c) of Ref. \cite{chisq}).  The number of
O phases increases as a function of $s$.  For $s=1$, our results
suggest that in addition to O$_1$, there is also an O$_2$ above, and its
complex-conjugate (c.c.) O$_2^*$ phase below, the FM phase.  There is no
definite indication of any other O phases. For $s=3/2$, there is an O$_1$
phase or a c.c. pair of O$_1$ and O$_1^*$ phases to the left of the FM phase,
and c.c. O$_2$ and O$_2^*$ phases to the upper and lower left of the FM
phase.  There may be other O phases also.  For $s=2$, there appear to be
several c.c. pairs of O phases in addition to the definite O$_1$ phase.

   Fourth, one may study special points separating CT phases.  Because of the
$u_s \to 1/u_s$ symmetry, and the consequent manner in which the phases group
themselves around the unit circle $|u_s|=1$, it follows that if two such
phases are contiguous at a single point, then this
point lies on this unit circle.  An exactly known example is the case
$u_{1/2}=-1$ for $s=1/2$.  From our $s=1$ zeros, it is
likely that one such point $u_{1;12}$
separates the O$_1$ and O$_2$ phase (so that also $u_{1;12}^*$ separates the
O$_1$ and O$_2^*$ phases); if this is the case, then it is likely that
$u_{1;12}=e^{2\pi i/3}$. For $s=3/2$, we infer likely intersection points of
phase boundaries at $u_{3/2;a}=i$ and
$u_{3/2;b}=e^{4\pi i/5} =  (1/4)(-\sqrt{5}-1+ i\sqrt{2}\sqrt{5-\sqrt{5}})
\simeq -0.809+0.588i$
together with their c.c.'s. For $s=2$, we infer likely intersection points at
$u_{2;a}=e^{\pi i/3}$,
$u_{2;b}=e^{2\pi i/5}=(1/4)(\ -1 +\sqrt{5}+
i\sqrt{2} \sqrt{5+\sqrt{5}}) \simeq 0.309 + 0.951 i$,
and $u_{2;c}=e^{4\pi i/5}$, together with their c.c's. There may be other such
points for $s=2$.

   Fifth, in addition to the zeros that lie on boundaries which, in the
thermodynamic limit, would separate CT phases, there are also zeros which
lie on arcs which
do not separate phases, but rather protrude into, and end in, the FM and AFM
phases.  As our exact solution of the CT phase diagram for the 1D spin $s$
Ising model \cite{is1d} shows, such arcs can represent the degeneracy, on a
finite curve, of the leading and next-to-leading eigenvalues of ${\cal T}$.
For $s=1/2$, there are no such protruding arcs for the square lattice, although
exact results show that they do occur for other 2D lattices (triangular
(in FM), honeycomb (PM), kagom\'e (PM), 3-12 lattice (PM) \cite{cmo}).
We denote the number of arcs protruding into the phase $ph$ (and terminating
in endpoints (e)) as $N_{e,ph}$. If these arcs occur at complex values of
$u_s$, then the $u_s \to u_s^*$ symmetry implies that they must occur as c.c.
pairs, so $N_{e,FM}$ is even.  For all of the $s$ values that we have studied
(on the square lattice), the arcs do occur at complex values of $u_s$, which
leads one to infer that $N_{e,FM}$ is even on the square for all $s$ (this will
follow from our conjecture (\ref{nefm}).\footnote{This inference does not
apply to other lattices; indeed, one knows of cases where $N_{e,FM}$ is odd:
e.g., exact results show that $N_{e,FM}=1$ for $s=1/2$ on the triangular
lattice.} Independent of this, for any bipartite lattice the $u_s \to 1/u_s$
symmetry implies that $N_{e,FM}=N_{e,AFM}$.

  For $s=1$ (Fig. 1(a)), we find an arc protruding in a ``NW-SE'' direction
into the FM phase and a corresponding arc protruding in a SE-NW direction
into the AFM phase, together with their complex conjugates, so $N_{e,FM}=2$
for $s=1$. For $s=3/2$ (Fig. 1(b)), we find three c.c. pairs of arcs protruding
into the FM phase, so $N_{e,FM}=6$.  For $s=2$ (Fig. 1(c)), we infer seven
such c.c. pairs of arcs, so $N_{e,FM} = 14$.
Our results are summarised in Table 1.\footnote{We add the usual caution that
analyses on finite lattices up to a given size can never rigorously
exclude the possibility that there are changes in the arcs as one gets
closer to the thermodynamic limit.}
For $s \ge 3/2$ we list the positions of the last zeros
to an accuracy of 0.01 in real and imaginary parts because this is typical of
the differences between the locations of these zeros for different lattice
sizes and BC's. The values of $|(u_s)_{e,Z}|$ are calculated using these
positions to their full accuracy for the lattices in Fig. 1.
The values of $(u_s)_c$ used for the last column are from Refs.
\cite{egj,jge}.

\begin{table}
\begin{center}
\begin{tabular}{|c|c|c|c|} \hline \hline \\
$s$ & $N_{e,FM}$ & $(u_s)_{e,Z}$ & $\frac{|(u_s)_{e,Z}|}{(u_s)_c}$ \\
 \hline \hline
1/2 & 0 & $-$ & $-$ \\ \hline
1   & 2 & $-0.305 \pm 0.381i$ & 0.874 \\ \hline
3/2 & 6 & $0.09 \pm 0.65i$ & 0.88 \\
    &   & $-0.07 \pm 0.72i$ & 0.98 \\
    &   & $-0.54 \pm 0.34i$ & 0.85 \\ \hline
2 &  14 & $0.38 \pm 0.65i$ & 0.90 \\
  &    & $0.30 \pm 0.73i$ & 0.95 \\
  &    & $0.21 \pm 0.80i$ & 1.0 \\
  &    & $-0.23 \pm 0.69i$ & 0.87 \\
  &    & $-0.40 \pm 0.69i$ & 0.965 \\
  &    & $-0.68 \pm 0.49i$ & 1.0 \\
  &    & $-0.655 \pm 0.29i$ & 0.86  \\ \hline \hline
\end{tabular}
\end{center}

\caption{Endpoints of arcs protruding into the FM phase.  Entries in column
denoted  $(u_s)_{e,Z}$ are the positions of the last zeros on the arcs in
Figs. 1(a)-(c).}
\label{table1}
\end{table}

   From previous studies of the $s=1/2$ Ising model for lattices in $d=2,3$,
we have been led to formulate several general conjectures for CT
singularities in $M$ and the susceptibility $\bar\chi$.
One of these is Conj. 1: For the zero-field $s=1/2$ Ising model on an arbitrary
(regular) lattice with even coordination number $q$, $M$ diverges at a point
$z=e^{-2K}$ if and only if $z$ is an endpoint of an arc (of non-analyticities
of $f$, as discussed above) protruding into the FM phase.  A corollary of
Conj. 1 is that on a bipartite lattice, for each such endpoint there is
another with $K \to -K$ at which the staggered magnetisation $M_{st}$
diverges.\footnote{We record a related conjecture here \cite{cmo}:
Conj. 2: For the $s=1/2$ Ising model on an arbitrary (regular) lattice with odd
coordination number $q$, in addition to the divergences at arc endpoints
specified by Conj. 1, $M$ also diverges at the point $z=-1$ if and only if
this point can be reached from within the FM phase.}  In Ref. \cite{chitri}
we proved (as Theorem 1 in the first paper) that if $M$ diverges at a given
point $z=e^{-2K}$, so does $\bar\chi$, and thus, for a bipartite lattice,
if $M_{st}$ diverges, so does the staggered susceptibility,
$\chi^{(a)}$.\footnote{Our study of the exactly solvable $s=1/2$ Ising model
with $\beta H = i \pi/2$ in Ref. \cite{ih} suggests that it is possible to
generalise Conj. 1 to a class of nonzero external fields.}  Since there is no
obvious reason why this conjecture should be limited to $s=1/2$, it is natural
to consider the following generalisation:
Conj. 1s: For the zero-field Ising model with arbitrary $s$
on an arbitrary (regular) lattice with even $q$, $M$
diverges at a point $u_s$ if and only if $u_s$ is an endpoint of
an arc of non-analyticities of $f$ protruding into the FM phase.
A corollary of Conj. 1s is that on a bipartite lattice, to each such point,
there corresponds a point $1/u_s$ in the AFM phase, and $M_{st}$ diverges at
a point $1/u_s$ in the AFM phase if and only if $1/u_s$ is the endpoint of
an arc protruding into the AFM phase (the image, under $u_s \to 1/u_s$, of
the arc in the FM phase).

   Combining our determination of the positions of the endpoints of arcs
protruding into the FM phase with the results of low-temperature series
analyses in Refs. \cite{egj} ($s=1$) and \cite{jge} $1 \le s \le 3$, we find
that for the values $s=1$, 3/2, and 2 where the comparison can be made, there
is excellent support for Conjecture 1s.  Of course, the position of the last
zero on an arc calculated for a finite lattice will not, in general, be
precisely the same as the endpoint in the thermodynamic limit, but we expect
it to be close for the lattices of the sizes that we are using.  For $s=1$,
Ref. \cite{egj} found CT divergences in $M$ at the c.c points
$u_1 = -0.30196(3) \pm 0.37875(5)i$, which match nicely with the arc
endpoints we have found for this case.  As a consequence of our
Theorem 1 in Ref. \cite{chitri}, these divergences in $M$ automatically imply
divergences in $\bar\chi$, and these were also observed in Ref. \cite{egj}.
For $s=3/2$ and 2, all of the divergences in $M$ again match the locations of
the arc endpoints which we have found \cite{jge}.  For example, for $s=3/2$,
preliminary results give \cite{jge} $u_{3/2}=0.0948(1) \pm 0.6410(2)i$ and
$-0.5291(2) \pm 0.3379(2)i$ for the two such singularities closest to the
origin, in agreement with our values in Table 1, with a similar match for
the third.  These comparisons indicate that the last zero on an arc for
the finite lattices which we have used is  a distance of about 0.01 farther
away from the origin than the singularity positions from series analyses
\cite{jge,egj}, which presumably reflect the thermodynamic limit.  (For $s=1$,
the difference in distances is less, about 0.004.) We infer that the last zeros
on the arcs give values of $\rho = |u_s|/(u_s)_c$ which are larger than
the exact values by roughly 1 \%.

 These results differ with an old conjecture \cite{fg} that for the spin $s$
Ising model the number $N_{cs,\rho < 1}$ of complex-temperature singularities
within the disk $\rho < 1$ is $N_{cs,\rho<1} = qs-2$, where $q$ is the
coordination number of the lattice. Here, this would yield
$N_{cs,\rho<1}=4s-2=2$, 4, and 6 for $s=1$, 3/2, and 2, rather than the
respective values 2, 6, and $\ge 10$ which we find with $\rho < 1$.

 Combining our present calculation of partition function zeros with an exact
determination of the CT phase structure of the 1D spin $s$ Ising model
\cite{is1d}, we make the following conjecture: For the spin $s$ Ising model
on the square lattice, the number of endpoints of arcs protruding into the CT
FM (equivalently, CT AFM phase) is
\beq
N_{e,FM}=N_{e,AFM}=4[s^2]-2
\label{nefm}
\eeq
for $s \ge 1$, where $[x]$ denotes the integral part of $x$. Combining this
with the exactly known result for $s=1/2$, the general formula reads
$N_{e,FM}=N_{e,AFM}=\max \{ \ 0, 4[s^2]-2 \}$.

We next discuss the $s$ dependence of the PM phase.   In general, for
a fixed lattice and temperature, an increase in $s$ has the effect of allowing
more randomness, and hence increasing the disorder in the model. This is
reflected in the well-known fact that $K_c$ is a monotonically increasing
($e^{-K_c}$ is a monotonically decreasing) function of $s$
\cite{fg,fg1,crit,jge}.  Thus, the disordered,
PM phase increases in extent as $s$ increases.  However, although $K_c$
increases as a function of $s$, the ratio $K_c/s^2$ which occurs in the
variable $u_s$ {\it decreases}, approaching zero as $s \to \infty$.
Indeed, as $s \to \infty$, $K_c$ approaches a finite limiting value,
$K_c(s) \nearrow K_c(s=\infty)$.  This follows since in
this $s \to \infty$ limit (renormalising $Z$ so that each
spin summation is normalised to measure 1 rather than $2s+1$), one obtains the
continuous-spin Ising model, with $Z=(\prod_n \int_{-1}^{1} (d\bar s_n/2))
\exp ( K \sum_{<nn'>}\bar s_n \bar s_{n'})$.  For $d > 1$
this model has a transition at a finite value of $K$ \cite{rbg}, viz.,
$\lim_{s \to \infty} K_c(s)$.  Since $K_c/s^2 \to const./s^2 \to 0$ as
$s \to \infty$, it follows that
$(u_s)_c \nearrow 1$ as $s \to \infty$.  Hence, in terms of
$u_s$, the physical PM phase appears to decrease in size as $s$ increases
and, in the limit as $s \to \infty$, to contract to a set of measure zero.
This has also been noted in Ref. \cite{jge}, where it was suggested that as $s
\to \infty$, both the physical critical point and the CT
singularities in $u_s$ will converge to the unit circle $|u_s|=1$. In the case
of the physical critical point, one should note, however, that this is really
due to the definition of $u_s$; in fact, in a fixed variable such as $K$ or
$v$, the physical PM phase actually increases in extent, reaching a limiting
size, as $s \to \infty$. From our plots of zeros, we conclude more generally
that not just the PM
phase but the O phases also cluster more closely around the unit circle
$|u_s|=1$ as $s$ increases.  We can thus state a stronger form of the
conjecture of Ref. \cite{jge}: As $s \to \infty$, the complex-temperature phase
boundaries will approach this unit circle, and in this limit, one will be
left with only the FM phase for $|u_\infty| < 1$ and the AFM phase for
$|u_\infty| > 1$.

This research was supported in part by the NSF grant PHY-93-09888.  We thank
Professor Tony Guttmann for kindly sending us a copy of Ref. \cite{jge} in
advance of publication and for interesting discussions of that work.

\vspace{6mm}

\begin{center}

{\bf Figure Captions}

\end{center}

 Fig. 1. Plots of zeros of $Z$ for the Ising model on the square lattice, as
functions of $u_s=e^{-K/s^2}$, for (a) $s=1$, lattice size $8 \times 10$,
PBC; (b) $s=3/2$, lattice size $6 \times 6$, PBC; (c) $s=2$, lattice
$5 \times (38/5)$, CHBC. See text for notation.
\vfill
\eject
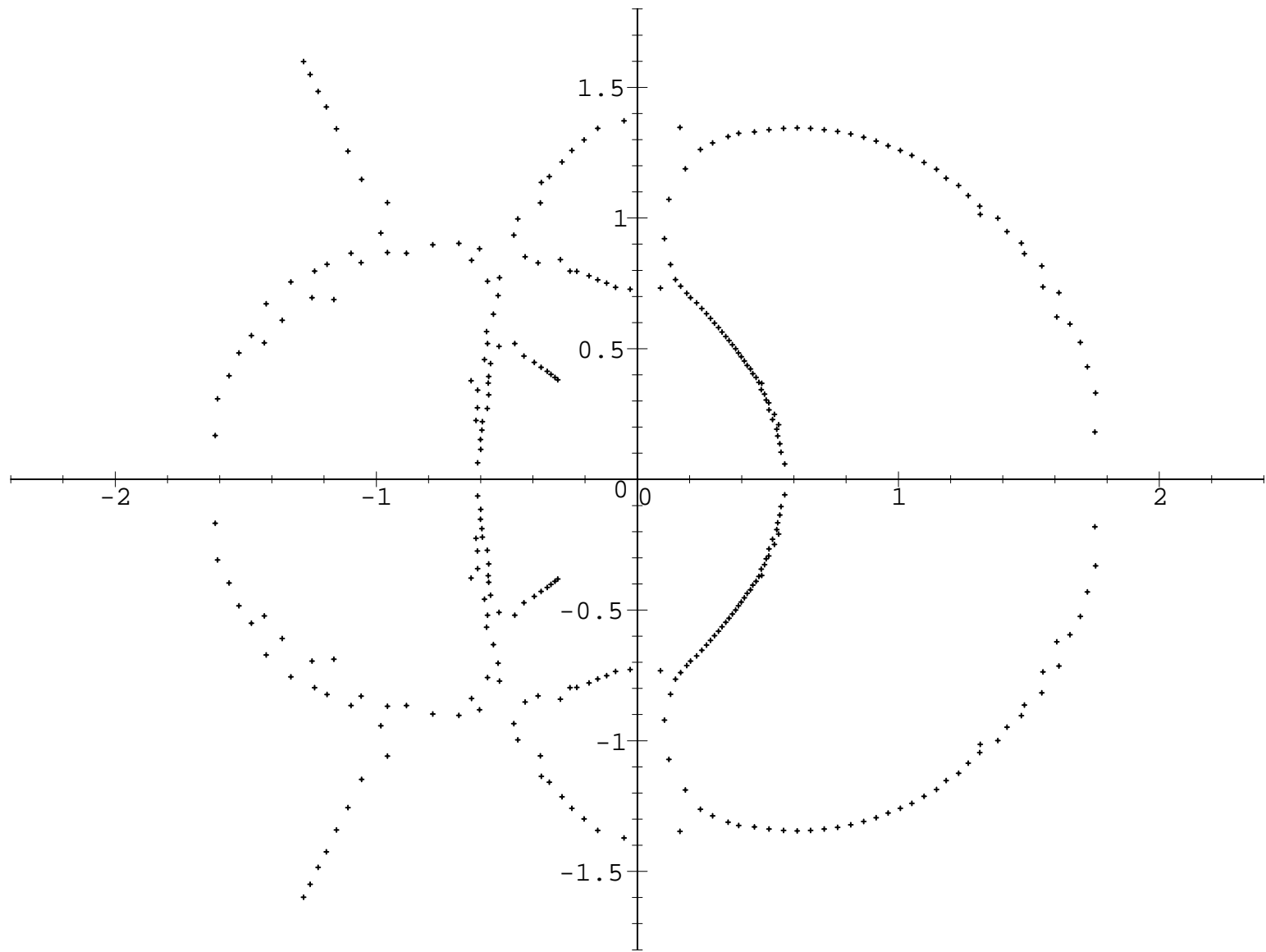
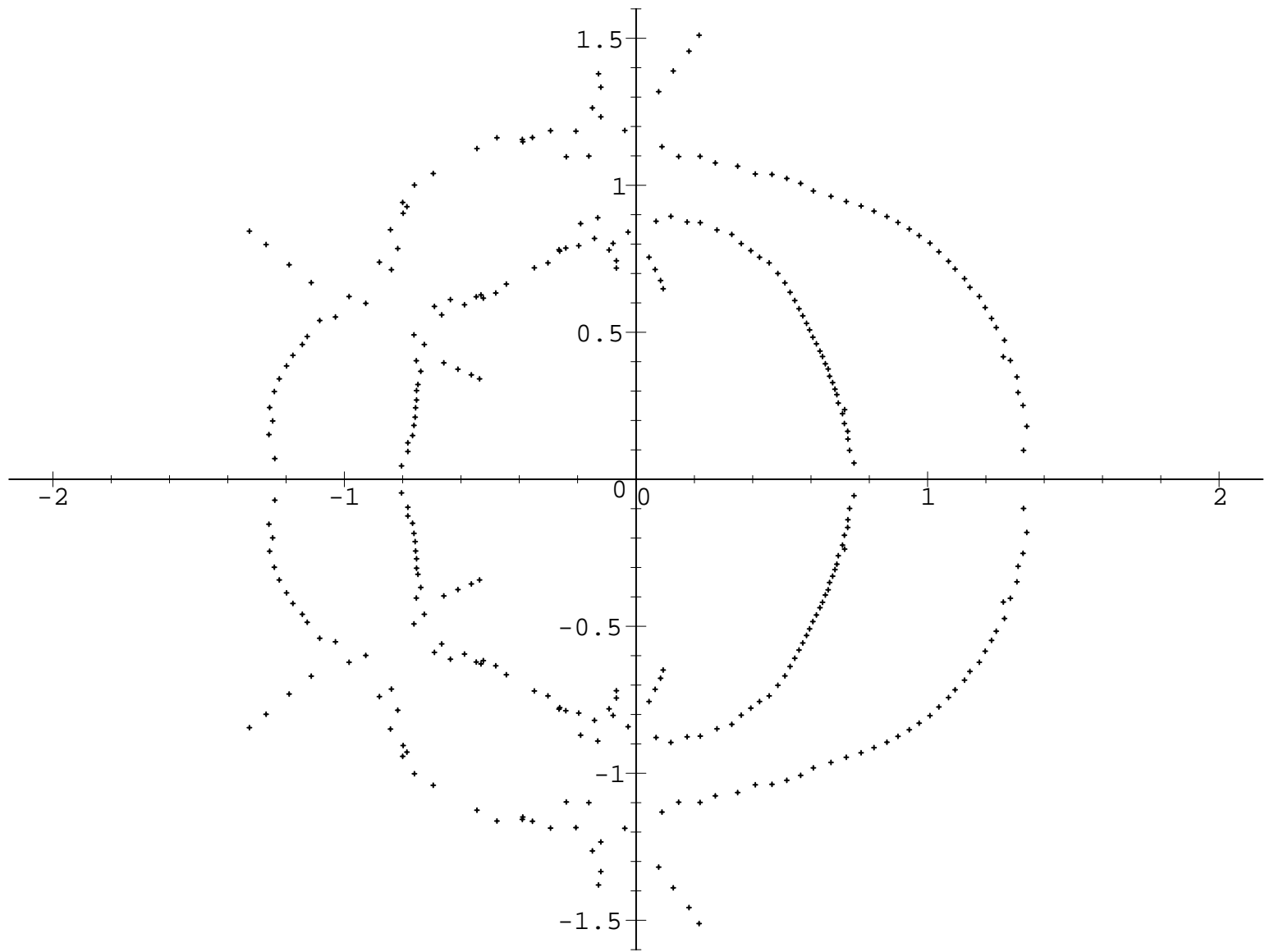
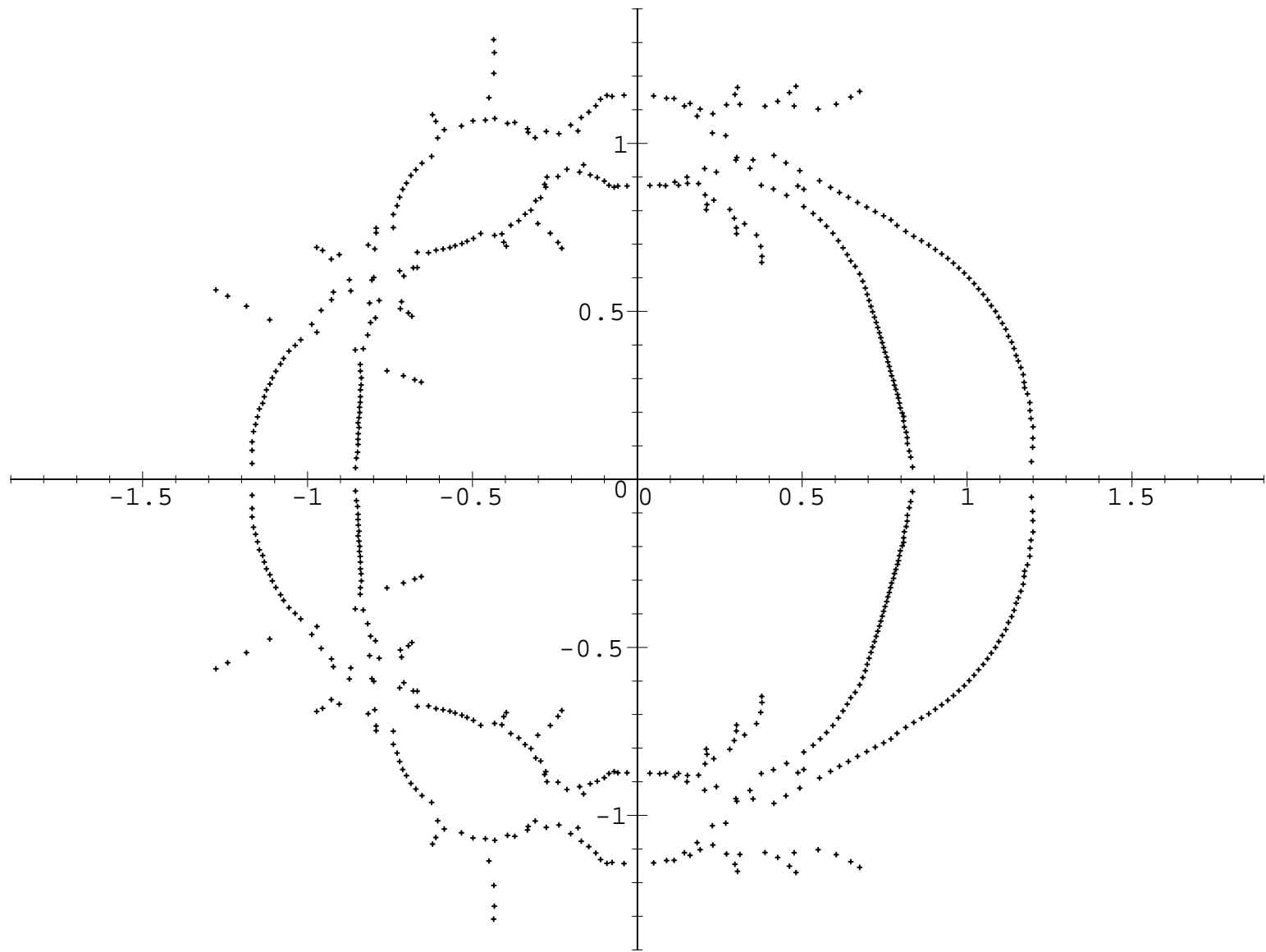
\end{document}